\pdfoutput=1
\RequirePackage{fix-cm}
\documentclass[british]{article}

\usepackage{spconf} 

\usepackage[utf8]{inputenc}
\usepackage[T1]{fontenc}
\usepackage[british]{babel}

\usepackage{pifont}

\usepackage[pdftex]{graphicx}
\usepackage[table,xcdraw]{xcolor}
\graphicspath{{./figures/}}
\DeclareGraphicsExtensions{.pdf,.png,.jpeg,.jpg}

\usepackage[cmex10]{amsmath} 
\usepackage{amssymb}
\usepackage{bm} 
\usepackage{esint} 
\usepackage{commath} 
\usepackage{units} 

\usepackage{subcaption} 

\usepackage{booktabs} 
\usepackage{array} 

\usepackage{multirow} 
\usepackage{rotating} 

\usepackage{enumitem} 
\setlist{nolistsep} 

\usepackage[unicode=true,pdfusetitle,%
 pdfauthor={Shivam Mehta, Éva Székely, Jonas Beskow, Gustav Eje Henter},%
 pdfkeywords={seq2seq, attention, HMMs, duration modelling, acoustic modelling},%
 bookmarks=true,bookmarksnumbered=true,bookmarksopen=false,%
 breaklinks=true,pdfborder={0 0 0},backref=false,colorlinks=true,citecolor=blue]{hyperref}

\newcommand{\bb}[1]{\boldsymbol{#1}}
\newcommand{\seq}[1]{#1} 

\renewcommand{\mid}{\,\ifnum\currentgrouptype=16 \middle\fi|\,}

\newcommand{\x}{\boldsymbol{x}}
\newcommand{\xseq}{\seq{\x}}
\newcommand{\btheta}{\bb{\theta}}

\newcommand{\webpage}{https://shivammehta007.github.io/Neural-HMM/}

\newcommand{\tc}{T2}
\newcommand{\tcp}{T2$+$P}
\newcommand{\tcnop}{T2$-$P}
\newcommand{\nh}{NH}
\newcommand{\nhtwo}{NH2}
\newcommand{\nhone}{NH1}

\newcommand{\topdecoder}{OutputNet}

\ninept 

\let\oldmarginpar\marginpar
\renewcommand\marginpar[1]{\-\oldmarginpar[\raggedleft\footnotesize #1]%
{\raggedright\footnotesize #1}}

\title{Neural HMMs are all you need (for high-quality attention-free TTS)}

\name{Shivam Mehta, {\'E}va Sz{\'e}kely, Jonas Beskow, Gustav Eje Henter\thanks{This work was partially supported by the Wallenberg AI, Autonomous Systems and Software Program (WASP) funded by the Knut and Alice Wallenberg Foundation.}}
\address{Division of Speech, Music and Hearing, KTH Royal Institute of Technology, Stockholm, Sweden}

\begin{document}
\maketitle
\begin{abstract}
Neural sequence-to-sequence TTS has achieved significantly better output quality than statistical speech synthesis using HMMs. However, neural TTS is generally not probabilistic and uses non-monotonic attention. Attention failures increase training time and can make synthesis babble incoherently. This paper describes how the old and new paradigms can be combined to obtain the advantages of both worlds, by replacing attention in neural TTS with an autoregressive left-right no-skip hidden Markov model defined by a neural network. Based on this proposal, we modify Tacotron 2 to obtain an HMM-based neural TTS model with monotonic alignment, trained to maximise the full sequence likelihood without approximation. We also describe how to combine ideas from classical and contemporary TTS for best results. The resulting example system is smaller and simpler than Tacotron 2, and learns to speak with fewer iterations and less data, whilst achieving comparable naturalness prior to the post-net. Our approach also allows easy control over speaking rate.
\end{abstract}
\section{Introduction}
\label{sec:intro}
Text-to-speech (TTS) technology has advanced tremendously in the last decade, and output speech quality has seen a number of step changes as the field evolved.
Statistical parametric speech synthesis (SPSS) based on hidden Markov models (HMMs)
\cite{zen2009statistical}, 
has now largely been supplanted by neural TTS \cite{tan2021survey}.
Waveform-level deep learning
greatly improved segmental quality over signal-processing based vocoders,
while sequence-to-sequence models with attention, e.g., \cite{wang2017tacotron}, demonstrated greatly improved prosody.
Combined, as in Tacotron 2 \cite{shen2018natural}, these innovations produce synthetic speech whose naturalness sometimes rivals that of recorded speech.

However, not all aspects of TTS systems have improved along the way.
The integration of deep learning with positional features into HMM-based TTS increased naturalness \cite{watts2016hmms}, but sacrificed the ability to learn to speak and align simultaneously, instead requiring an external forced aligner.
Attention-based neural TTS systems \cite{wang2017tacotron} reintroduced the ability to learn to align, but are not grounded in probability and require more data and time to start speaking.
Furthermore,
their non-monotonic attention mechanisms 
do not enforce a consistent ordering of speech sounds.
As a result, synthesis is susceptible to skipping and stuttering artefacts (as seen in \cite{watts2019where}), and may break down catastrophically, resulting in unintelligible gibberish.

In this article, we 1) make the case that HMM-based and neural TTS approaches can be combined to gain the benefits of both worlds.
We 2) support this claim by describing a neural TTS architecture based on Tacotron 2, but with the attention mechanism replaced by a Markovian hidden state, to obtain a fully probabilistic, joint model of durations and acoustics.
The model development leverages design principles from both HMM-based and sequence-to-sequence TTS.
Experiments show that the model gives
a speech quality on par with that of a comparable Tacotron 2 model, and produces intelligible speech already after 1k updates, a 15-fold improvement on Tacotron 2.
Unlike standard Tacotron 2, it also allows control over speaking rate.
For audio examples and code, please \href{\webpage}{see our demo webpage}.

\section{Background}
\label{sec:background}
The starting point of this work is \cite{watts2019where}, which identified four key differences between HMM-based SPSS
and sequence-to-sequence attention-based TTS
that had a notable impact on output quality:
\begin{enumerate}
\item \label{it:vocoder} Neural vocoder with mel-spectrogram inputs
\item \label{it:frontend} Learned front-end (the encoder)
\item \label{it:ar} Acoustic feedback (autroregression)
\item \label{it:attention} Attention instead of HMM-based alignment
\end{enumerate}
Among these, items \ref{it:vocoder}--\ref{it:ar} led to improved speech quality, whereas attention
sometimes made the output significantly worse.
This paper incorporates
aspects \ref{it:vocoder}--\ref{it:ar} into a TTS system that leverages 
neural HMMs \cite{tran2016unsupervised,yu2016online} rather than attention for sequence-to-sequence modelling.
Sec.\ \ref{ssec:upgrading}, below, describes how to add aspects \ref{it:vocoder}--\ref{it:ar} to HMMs based on prior work, with attention (aspect \ref{it:attention}) discussed in Sec.\ \ref{ssec:attention}.

\subsection{Adding neural TTS aspects to HMM-based TTS}
\label{ssec:upgrading}
For aspect \ref{it:vocoder}, high-quality neural vocoders are now available off the shelf.
Furthermore, most of these use spectral features as input.
This helps avoid flat intonation caused by explicit averaging over pitch contours, commonly seen in systems that use a separate $f$0 feature to parameterise speech \cite{watts2019where}.
However, nothing prevents HMM-based TTS from using mel-spectrogram features and neural vocoders: this is just a straightforward change of acoustic features, and the HMM-based approach described in this paper uses this setup.

Another factor in the improved prosody is item \ref{it:frontend}, the learned front-end (i.e., the encoder).
Again, there is nothing that prevents using this idea in a system that leverages HMMs.
The HMM-based systems we introduce all use the same encoder architecture as Tacotron 2 \cite{shen2018natural} with no additional linguistic features added.

The situation for item \ref{it:ar}, autoregression (AR), is again similar, in that AR and HMMs are not mutually exclusive.
Acoustic models in HMM-based TTS systems 
benefit from using positional and durational information \cite{wu2016merlin,watts2016hmms}, that increases granularity by enabling the statistics of each generated frame to be different,
together with dynamic features \cite{tokuda2000speech}
to promote continuity across time.
However, positional and durational features violate the Markov assumption (e.g., they depend on the time spent in the current state), preventing realignment during TTS training.
In a model like Tacotron, positional information is instead mediated and continuity enforced by autoregression.
Since this
only involves dependencies on observed variables, it is possible to
devise autoregressive models that do not violate the Markov assumption, and linear autoregressive HMMs (AR-HMMs) \cite{rabiner1989tutorial}
have previously been explored in HMM-based SPSS \cite{shannon2013autoregressive,quillen2012autoregressive,wang2017autoregressive}.
In this paper, we describe HMMs that, like Tacotron, use stronger, nonlinear AR models defined by a neural network.

\subsection{Attention in TTS}
\label{ssec:attention}
In a typical sequence-to-sequence based TTS system, the attention mechanism is responsible for duration modelling and for learning to align input symbols with output frames during training.
Watts et al.\ \cite{watts2019where} found that the use of neural attention did not necessarily benefit TTS,
and more suitable TTS attention mechanisms have recently been a focus of intense research.
Only some of the relevant work can be surveyed here; please see \cite{tan2021survey} for additional references. He et al.\ \cite{he2019robust} emphasised that TTS alignments should be \emph{local} (each output frame is associated with a single input symbol), \emph{monotonic} (never move backwards), and \emph{complete} (not skip any speech sounds).
HMMs are local by design, while the two other concepts map directly onto the classes of \emph{left-right} and \emph{no-skip} HMMs.
Most neural TTS attention mechanisms do not satisfy these requirements \cite{he2019robust,tan2021survey}.


Many systems that do satisfy all three criteria
rely on external tools for input-output alignment to obtain duration data (see \cite{tan2021survey} for a list), and do not jointly learn to speak and align, unlike regular HMMs or Tacotron 1/2.
However, some proposals
do learn to speak and align without external tools,
mostly (e.g., \cite{yasuda2019initial,zeng2020aligntts,kim2020glow,shen2020non,yasuda2021end,nankaku2021neural,fujimoto2022autoregressive}) by introducing duration models into neural TTS, which will be our focus here.
%
Many of these models only optimise a lower bound on the sequence likelihood, either due to the use of variational methods (e.g., Non-Attentive Tacotron \cite{shen2020non} and the VQ-VAEs in \cite{yasuda2021end}) or by not marginalising over all possible alignments (Glow-TTS \cite{kim2020glow}).
By using a mean squared error (MSE) duration loss, Glow-TTS also implicitly treats the positive, integer-valued durations (frame counts) as outcomes from a Gaussian distribution on the real line, which violates probabilistic assumptions.
Our proposal avoids these issues.

AlignTTS \cite{zeng2020aligntts} is more similar to an HMM and uses a variant of the HMM forward recursions \cite{rabiner1989tutorial}, but requires a complex, four-stage training procedure that culminates in training a separate, non-probabilistic duration predictor that is used at synthesis time.
AlignTTS is also parallel, while our proposal is autoregressive.


The constant-per-state transition probability of regular HMMs implicitly describes a
geometric duration distribution, which is a poor fit for natural speech
\cite{zen2004hidden,ronanki2016median}.
A solution to this in SPSS was to introduce explicit duration modelling through \emph{hidden semi-Markov models} (HSMMs) \cite{zen2004hidden}.
These sacrifice the Markovian property to describe more general duration distributions, by letting transition probabilities depend on the time spent in the current state.
Independent work \cite{nankaku2021neural,fujimoto2022autoregressive} concurrent to ours proposes to integrate HSMMs into neural TTS, obtaining better results than Tacotron 2, but uses a variational approximation and again assumes a Gaussian distribution for the positive-integer frame durations.
In contrast, \cite{ronanki2016median} described how
arbitrary discrete
duration distributions can be parameterised implicitly via frame-dependent transition probabilities, and then predicted jointly with output frames in a single, joint model of durations and acoustics.
This paper 
combines this idea with autoregression acting as an indirect, ``acoustic memory'' of the time spent in a state, 
to obtain a fully probabilistic model with general discrete durations, that can be trained efficiently on the exact log-likelihood. 

The most similar work to ours is SSNT-TTS \cite{yasuda2019initial}, which essentially describes a neural HMM for TTS, albeit under another name.
We differ in applying an HMM perspective to the approach, in integrating more SPSS ideas to improve our system, in using a different duration-generation method, in demonstrating control over speaking rate, and in reporting better TTS quality, on par with Tacotron 2.

\section{Method}
\label{sec:method}
\begin{figure*}[!t]
  \centering
  \subcaptionbox{Nvidia Tacotron 2 implementation\label{sfig:tacotron}}{%
      \includegraphics[width=\columnwidth]{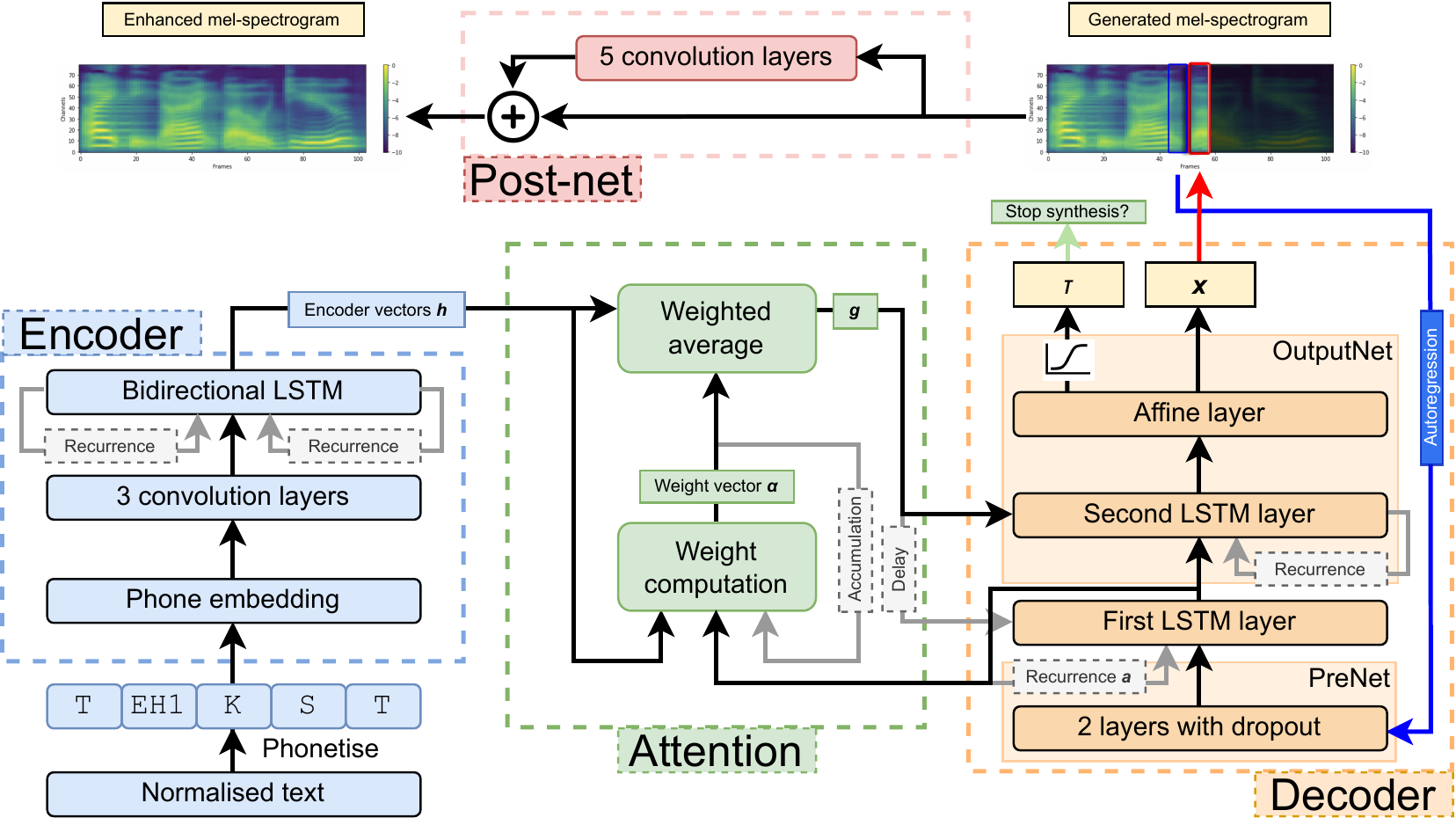}}%
  \hfill
  \subcaptionbox{Tacotron 2 architecture modified to use a neural HMM\label{sfig:neuralhmm}}{%
      \includegraphics[width=\columnwidth]{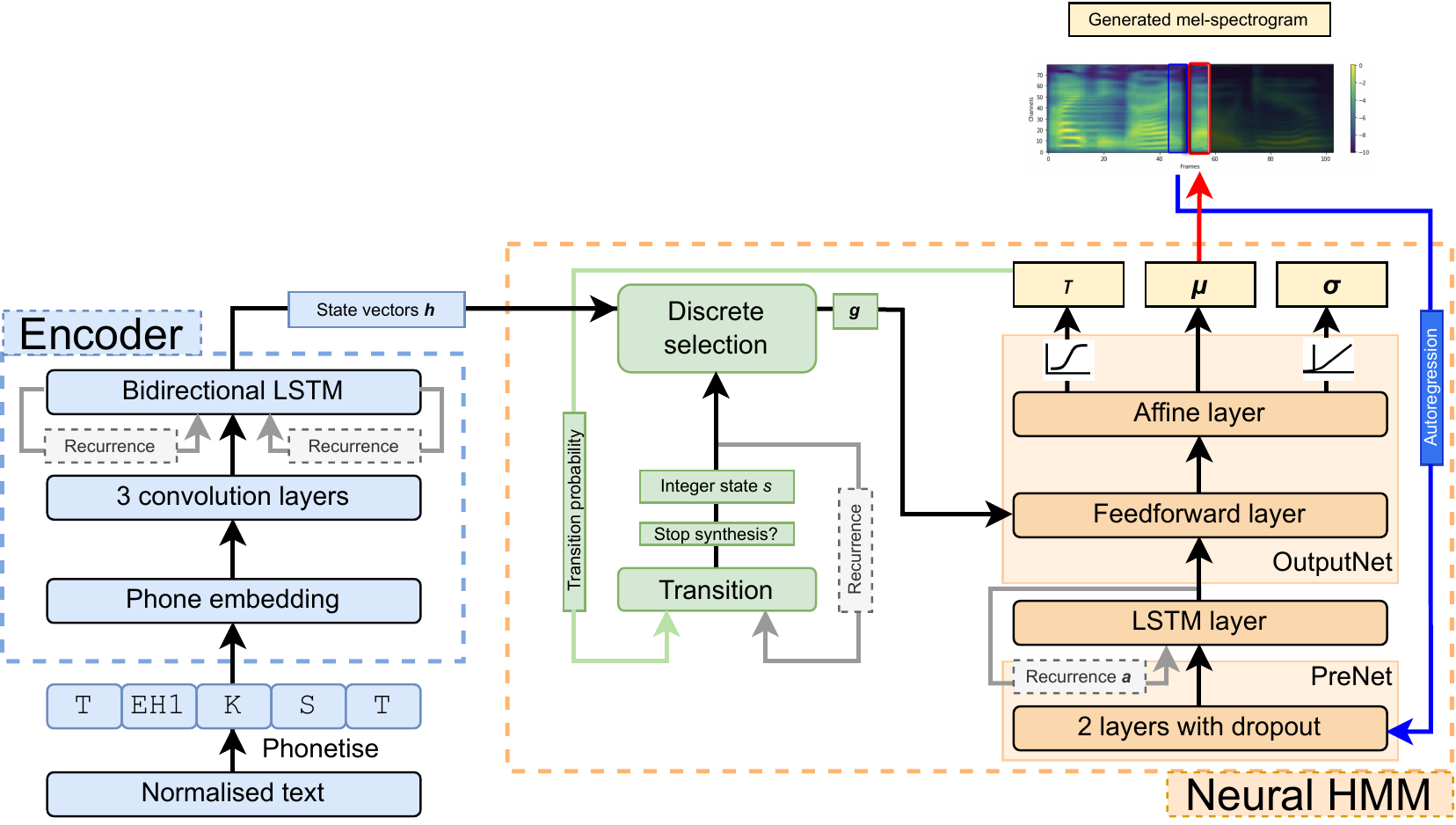}}%
  \caption{Synthesis-time architecture diagrams. Recurrences, delays, and the cumulative attention in Eq.\ \eqref{eq:decoder2} are drawn as grey arrows.}
  \label{fig:diagrams}
  \vspace{-1\baselineskip}
\end{figure*}

We now (in Sec.\ \ref{ssec:neuralhmm} and Fig.\ \ref{fig:diagrams}) describe the key modifications used to put HMMs into neural TTS such as Tacotron 2.
Sec.\ \ref{ssec:implementation} then describes how ideas and implementation aspects from classic HMM-based TTS can be adapted to further improve neural HMM TTS.

\subsection{Replacing attention with neural HMMs}
\label{ssec:neuralhmm}
The location-sensitive attention \cite{chorowski2015attention} used by Tacotron 2 is a function that uses information from previously generated acoustic frames $\xseq_{1:t-1}$ to 
select which encoder output vector(s) $\bb{h}_n$ to send to the decoder, to generate the next frame $\x_t$.
(We use bold font for vector-valued quantities and index input-sequence symbols by $n$ and output frames by $t$.)
The attention also has an internal state, in the form of previous attention weights $\alpha_{1:t-1,n}$.
Fig.\ \ref{sfig:tacotron} shows the procedure to generate one frame $t$ of output using Tacotron 2. It can be written as
\begin{align}
\bb{a}_t & = \mathrm{LSTM}(\mathrm{PreNet}(\x_{t-1}),\bb{g}_{t-1},\bb{a}_{t-1}) \label{eq:decoder1}\\
e_{t,n} & = \bb{\omega}^{\intercal} \tanh\big(\bb{W}\bb{a}_t + \bb{V}\bb{h}_n + \bb{U}(\bb{F}\ast\textstyle{\sum_{t'<t}}\alpha_{t',n}) + \bb{b}\big) \label{eq:decoder2}\\
\alpha_{t,n} & = \exp(e_{t,n}) \mathbin{/} \textstyle{\sum_{n'}} \exp(e_{t,n'}) \\
\bb{g}_t & = \textstyle{\sum_{n}} \alpha_{t,n} \bb{h}_n \qquad{}
(\x_t,\tau_t) = \mathrm{\topdecoder{}}(\bb{g}_t,\bb{a}_t) \label{eq:decoderlast} 
\text{.}
\end{align}
Here, $\bb{a}_{t-1}$ represents the hidden and cell state variables of the first decoder LSTM, $\mathrm{\topdecoder{}}$ is the upper part of the decoder in Fig.\ \ref{sfig:tacotron} (which contains a second LSTM), while $\tau_t\in[0,\,1]$ is the \emph{stop token}.
The latter is an estimate of the probability that the current frame is the last in the utterance, terminating synthesis if $\tau_t>0.5$.

To swap in neural HMMs, we remove
the dependence on $\bb{g}_{t-1}$ from Eq.\ \eqref{eq:decoder1}, and replace attention by a probabilistic $\mathrm{\topdecoder{}}$ that uses $\bb{a}_{t}$ and the HMM \emph{state} $s_t\in\{1,\,\ldots,\,N\}$ to estimate the distribution of
frame $\x_t$, by outputting the parameters $\btheta_t$ of an HMM emission distribution $\bb{o}(\btheta)$.
The stop token becomes a \emph{transition probability} $\tau_t\in[0,\,1]$ for $s_t$, 
with $s_1=1$.
Eqs.\ \eqref{eq:decoder2}--\eqref{eq:decoderlast} then become
\begin{align}
\bb{g}_t & = \bb{h}_{s_t} \label{eq:neural1} \qquad{}
(\btheta_t,\tau_t) = \mathrm{\topdecoder{}}(\bb{g}_t,\bb{a}_t) \\
\x_t & \sim \bb{o}(\btheta_t) \qquad{}\;
s_{t+1} = s_{t} + \mathrm{Bernoulli}(\tau_t) \label{eq:neurallast}
\text{,}
\end{align}
where $\mathrm{Bernoulli}(p)$ is a binary random variable on $\{0,\,1\}$ that equals 1 with probability $p$.
The attention state variables $\alpha_{t,n}$ of Tacotron 2 have thus been replaced by a single, integer state variable $s_t$ that evolves stochastically based on $\tau_t$.
This transition probability depends on the $\bb{h}$-vector of the current state $s_t$ (through $\bb{g}_t$) and on the entire previous acoustics $\xseq_{1:t-1}$ (through $\bb{a}_t$), so it can be different for every frame $t$ even for the same state.
This can model arbitrary duration distributions \cite{ronanki2016median}.
$s_{t}>N$ terminates synthesis.

The end result is a left-right no-skip \emph{neural HMM}, an AR-HMM parameterised by the decoder network in Fig.\ \ref{sfig:neuralhmm}.
The encoder turns each input sequence
into a unique HMM, where each vector $\bb{h}_n$ represents a state.
Feeding this state vector and the AR input $\xseq_{1:t-1}$ into the decoder yields the HMM emission distribution $\bb{o}(\btheta_t)$ and next-state transition probability $\tau_t$ of state $n$ at time $t$.
Neural HMMs were first described concurrently by \cite{tran2016unsupervised} and \cite{yu2016online}, the latter under the name \emph{segment-to-segment neural transduction} (SSNT).

For the model to be a proper HMM satisfying the Markov property, $(\btheta_t,\,\tau_t)$ must not depend on anything other than the current state $s_t$ (through the state vector $\bb{g}_t$) and the past observations $\xseq_{1:t-1}$.
This necessitates an additional change to the Tacotron 2 architecture, namely removing the recurrence inside $\mathrm{\topdecoder{}}$ by changing its LSTM layer to a feedforward layer, since
an LSTM would propagate a dependence on past hidden states.
This change also substantially reduces the number of parameters in the model.

Finally, the full Tacotron 2 architecture contains a non-causal convolutional \emph{post-net} that enhances the initial AR-generated mel-spectrogram in a residual setup.
This
resembles post-filtering and global variance compensation \cite{toda2007speech} in classic SPSS.
Tacotron 2 training minimises the sum of the MSEs before and after the post-net.
However, the non-invertibility of the Tacotron post-net makes it incompatible with likelihood-based models like ours.
A post-net can be added, but must either be trained separately,
or be invertible like in \cite{kim2020glow}.
We leave this as future work,
and instead evaluate our proposal against Tacotron 2 output from both before and after the post-net.





\subsection{Practical considerations}
\label{ssec:implementation}
\textbf{Numerical stability:}
When working with HMMs, it is crucial for numerical precision to perform all computations in the logarithmic domain
using the ``log-sum-exp trick''.
Since zeroes in these computations map to $\ln0=-\infty$ in the log domain,
care must be taken
to avoid \texttt{NaN} gradients in deep-learning frameworks like PyTorch.

Like classic HMM-based TTS \cite{zen2009statistical}, we chose to use diagonal-covariance Gaussian emission distributions $\bb{o}(\bb{\mu},\,\bb{\sigma})$ in this work.
We also used softplus (not exponential) nonlinearities for $\bb{\sigma}$, with a non-zero minimum value (``variance flooring''), here clamped at 0.001, since this has been important in other generative models.

\textbf{Architecture enhancements:}
Tacotron 2 can represent intermediate states using soft attention, since the $\alpha_{t,n}$-values have many degrees of freedom.
Major HMM-based synthesisers instead use 5 sub-states per input phone and run at 200 fps \cite{zen2009statistical,wu2016merlin}.
Tacotron 2 runs at 80 fps, i.e., 40\% the framerate, hence we use 2 states per phone to get the same time resolution as these HMMs.
This is implemented by doubling the size of the decoder output layer and interpreting its output as two concatenated state vectors $\bb{h}$ for each phone.

Classic HMM-based TTS
includes a model of the dependencies between several adjacent frames to promote temporally smooth output \cite{zen2009statistical,shannon2013autoregressive,wu2016merlin}.
Although Tacotron 2 and the neural HMMs in this article only take the latest frame $x_{t-1}$ as AR input, the LSTM in Eq.\ \eqref{eq:decoder1} means they can remember information arbitrarily far back,
which is beneficial for modelling utterance-level prosody.
We also treat $\x_0$, the initial AR context (the ``go token''), as a learnable parameter.

\textbf{Initialisation:}
HMMs are often initialised using a \emph{flat start}, in which all states have the same statistics \cite{young2002htk}.
By zeroing out all weights in the decoder output layer but initialising other layers as normal, all states will have the same output (zero), but different and nonzero gradients, thus enabling learning \cite{zhang2019fixup}.
The last-layer bias values were chosen so that $\bb{\mu}=\bb{0}$ and $\bb{\sigma}=\bb{1}$ for every state at the start of training, to match the global statistics of our normalised data.

\textbf{Training:}
Neural HMM training \cite{tran2016unsupervised} is a hybrid of old and new:
We use the classic (scaled) forward algorithm \cite{rabiner1989tutorial} to compute the exact sequence log-likelihood, but then leverage backpropagation and automatic differentiation to optimise it using Adam.
These parts correspond to the E step and the M step of the (generalised) EM algorithm \cite{dempster1977maximum}, respectively.
Computations during training parallelise over the states but, like Tacotron 2, are sequential across time due to the temporal recurrences.

Maximum-likelihood estimation of linear AR-HMMs can lead to unstable models \cite{quillen2012autoregressive,shannon2013autoregressive}.
A similar problem exists for nonlinear, autoregressive neural TTS \cite{tan2021survey}.
Tacotron 2 works around this by adding dropout to the pre-net, and we retain that solution here.

\textbf{Synthesis:}
We can iteratively use the equations in Sec.\ \ref{ssec:neuralhmm}
and randomly sample new frames $\x_t\sim\bb{o}(\btheta_t)$.
However, HMM-based TTS generally benefits from deterministically generating typical output rather than random sampling \cite{henter2014measuring,henter2016robust}.
For acoustics, this is done by generating the most probable output sequence \cite{tokuda2000speech}, which is the same as the mean $\bb{\mu}_t$ when $\bb{o}(\btheta_t)$ is Gaussian.
By iteratively taking $\x_t=\bb{\mu}_t$ (red arrow in Fig.\ \ref{sfig:neuralhmm}), we obtain a greedy approximation of \cite{tokuda2000speech}.
This is closely related to Tacotron 2 output generation, since it is trained using the MSE, which is minimised by the mean $\mathbb{E}[\bb{X}_t]$.

SSNT-TTS found that randomly sampling transitions led to poor pause durations when synthesising \cite{yasuda2019initial}, and classic HMM-based systems typically base the time in each state on the mean duration of the state \cite{zen2004hidden}.
This mean is difficult to compute with duration distributions implicitly defined through transition probabilities $\tau_t$, as here.
We instead use the simple algorithm from \cite{ronanki2016median,henter2017nonparametric} for deterministic duration generation based on duration quantiles (e.g., the median rather than the mean).
A quantile threshold controls speaking rate, which can be adjusted on a per-state basis, unlike \cite{bae2020speaking}.
For the models evaluated in this paper, informal listening showed that deterministic generation of acoustics and durations both led to clear quality improvements; examples are provided on the webpage.


\section{Experiments}
\label{sec:experiments}
To validate our proposal and show
that neural HMMs provide notable advantages over attention in neural TTS,
we performed a number of experiments (including a subjective listening test) comparing TTS using neural HMMs to a maximally similar Tacotron 2 \cite{shen2018natural} system.
Synthetic speech examples from the different experiments can be found at \href{\webpage}{\webpage}.
\begin{figure}[!t]
  \centering
  \includegraphics[width=\columnwidth]{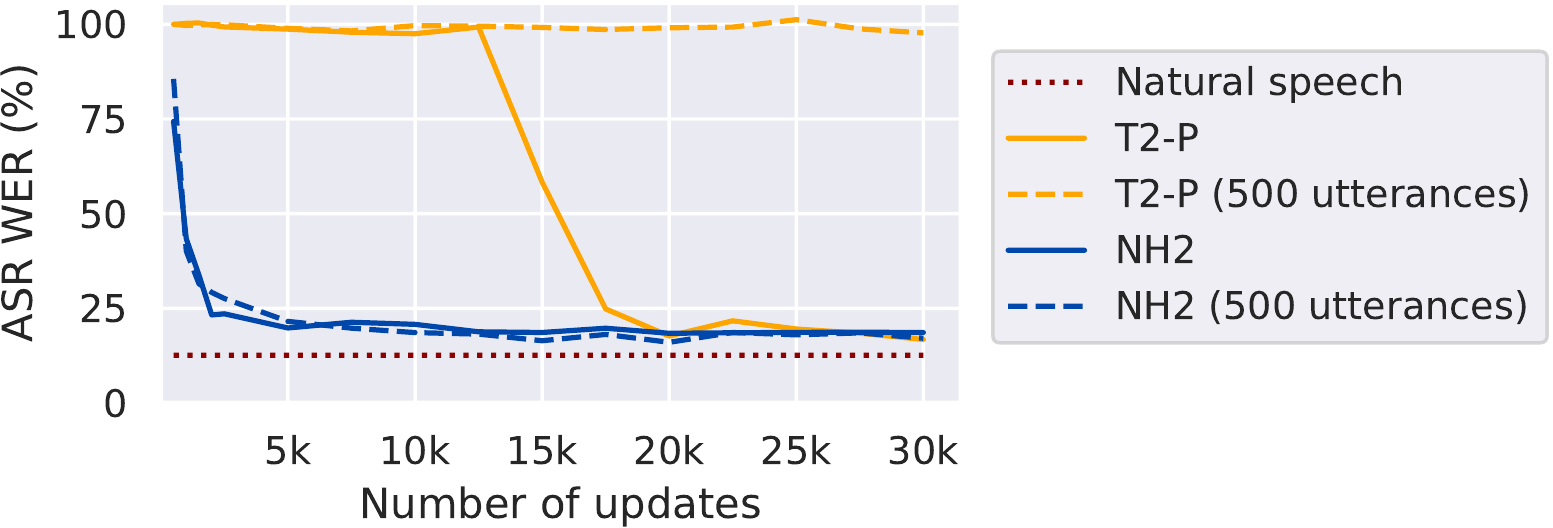}%
  \caption{Average utterance ASR WER of validation-set resynthesis.}
  \label{fig:wer}
  \vspace{-1\baselineskip}
\end{figure}

We based our systems on the widely used PyTorch
open-source Nvidia  implementation\footnote{\href{https://github.com/NVIDIA/tacotron2/}{https://github.com/NVIDIA/tacotron2/}} of Tacotron 2.
The systems were trained on the LJ Speech dataset\footnote{\href{https://keithito.com/LJ-Speech-Dataset/}{https://keithito.com/LJ-Speech-Dataset/}}, which contains utterances (normalised text and matching audio) adapted from free audiobooks read by a female speaker of US English.
We used the default train/val/test split in the repository, which designates
about 23 h of audio for training.
We likewise used the default text-processing, including the pronouncing dictionary (CMUdict), since this generally benefits neural TTS \cite{fong2019comparison}.
Output features were normalised to zero mean and unit variance over the training data, and waveforms were generated using the default, pre-trained v5 ``universal'' WaveGlow \cite{prenger2019waveglow} vocoder.\footnote{\href{https://github.com/NVIDIA/waveglow/}{https://github.com/NVIDIA/waveglow/}}

We trained three systems: one Tacotron 2 baseline (\tc{}) and two neural HMM systems, with either two (\textbf{\nhtwo{}}) or one (\textbf{\nhone{}}) state per phone.
We expect \nhtwo{} to perform the best, with \nhone{} functioning as an ablation.
All systems used the same architecture and hyperparameters (layer widths, learning rates, etc.) as the
repository defaults, except that the size of the decoder output vectors was doubled to 1024 in the two-state system, since the decoder output now represents two concatenated state vectors.
From the single Tacotron 2 baseline system, we synthesised two outputs:
\textbf{\tcp{}}, using the full mel-spectrogram output after the post-net, and \textbf{\tcnop{}}, using the initial mel-spectrogram prior to post-net enhancement, which is directly comparable to our neural HMMs.
Model sizes for the different setups are listed in Table \ref{tab:results}.
We see that both neural HMMs are significantly smaller than Tacotron 2, even if the post-net is removed.

Each system was trained for 30k mixed-precision updates on 7 GPUs using a batch size of 6.
It took approximately 14.5k updates for \tc{} to learn to
speak coherently, whereas \nhtwo{}
was intelligible after 1k updates.
Fig.\ \ref{fig:wer} graphs how the Google ASR word error rate (WER) of synthesising the 100 validation utterances evolves during training, including results from training on a small subset (500 utterances) of the data.
Audio of speech synthesised during training is also provided on our demo webpage.
We see that \nhtwo{} rapidly learns to speak intelligibly in both cases, much faster than Tacotron 2, which does not learn to speak at all on the smaller dataset.
Even after the WER stabilised, we could consistently reproduce the effect where Tacotron 2 (including the best pre-trained system made available by Nvidia) degenerates into unintelligible babbling on long and short sentences, with examples provided on our webpage.

Tacotron 2 applies pre-net dropout both during training and synthesis \cite{shen2018natural}, otherwise attention breaks down.
Our neural HMMs retained this dropout, since it improved the speech quality in informal listening.
Audio synthesised without it is provided on our webpage.

The distribution of phone durations in natural speech is skewed to the right.
The median of a skewed distribution lies between the mode and the mean, and median-based duration generation therefore often gives a faster-than-average speaking rate; cf.\ \cite{henter2016robust}.
Following the proposal in \cite{henter2017nonparametric}, the transition threshold of the deterministic duration-generation procedure was manually tuned to make the speaking rate of the \nh{} systems match \tc{}.
The resulting threshold-quantile values were 0.57 for \nhtwo{} and 0.45 for \nhone{}.
Our webpage provides examples of speech generated with different threshold quantiles, to demonstrate speaking-rate control at synthesis time.
\begin{table}[!t]
\centering
\begin{tabular}{@{}lcccc@{}}
\toprule 
Type & \multicolumn{2}{c}{Tacotron 2} & \multicolumn{2}{c}{Neural HMM}\\
Condition & \tcp{} & \tcnop{} & \nhtwo{} & \nhone{}\\
\midrule 
Size & 28.2M & 23.8M & 15.3M & 12.7M\\
MOS & 3.41$\pm$0.01 & 3.25$\pm$0.01 & 3.24$\pm$0.01 & 2.68$\pm$0.01\\
\bottomrule
\end{tabular}%
\caption{Models from the experiments, with number of parameters and mean opinion scores (with 95\% confidence intervals) for each.}
\label{tab:results}
\vspace{-0.8\baselineskip}
\end{table}


We conducted a subjective listening test to evaluate speech naturalness for the four conditions in Table \ref{tab:results}.
In the test, participants were presented with four parallel stimuli at a time, one from each condition (unlabelled and in random order), all speaking the same sentence.
Participants were asked to rate the naturalness of each stimulus on an integer scale from 1 (worst) to 5 (best), anchored using the classic MOS labels ``Bad'' through ``Excellent''.
Stimuli were drawn from a pool of 9 sets of Harvard sentences \cite{rothauser1969ieee}, which are sets of 10 sentences each, designed so that each set is approximately phonetically balanced.
All stimuli were loudness normalised to $-$20 dB LUFS following EBU R128 \cite{ebu2020loudness}.
We manually verified that no \tc{} stimuli exhibited babbling due to failed attention.

We used \href{https://prolific.co/}{Prolific} to recruit 30 test participants ages 21 through 70, all self-reported headphone-wearing native English speakers from UK, Ireland, USA, Canada, Australia, and New Zealand.
Each participant rated 3 randomly selected sets of 10 Harvard sentences, giving a grand total of 3600 ratings, 900 per condition.
A completed test took on average 17 minutes and was rewarded with 3.50 GBP.

The mean opinion scores (MOS) from the test are reported in Table \ref{tab:results}, together with 95\% confidence intervals based on a Gaussian approximation.
Pairwise $t$-tests find all conditions to be significantly different (with $p{<}10^{-3}$) except \nhtwo{} and \tcnop{} ($p{>}0.98$),
whose respective mean opinion scores differ by less than 0.002 before rounding.
We can conclude that the proposed neural HMM TTS (\nhtwo{}), despite being simpler and lighter, achieved a naturalness on par with the most comparable Tacotron 2 condition (\tcnop{}).
This was not achieved by SSNT-TTS \cite{yasuda2019initial}.
Neural HMMs were found to benefit from using two states per phone (\nhtwo{} vs.\ \nhone{}), whilst
Tacotron 2 improved from the use of a post-net (\tcp{} vs.\ \tcnop{}).


\section{Conclusion and future work}
\label{sec:conclusion}
We have described how classical and contemporary TTS paradigms can be combined to obtain fully probabilistic, attention-free sequence-to-sequence TTS based on neural HMMs.
Our example system is smaller than Tacotron 2, yet achieves comparable naturalness, learns to speak and align faster, needs less data, and does not babble.
To our knowledge, this is the first time an HMM-based system demonstrates a speech quality matching prior neural TTS.
The neural HMMs also permit easy control over the speaking rate of the synthetic speech.

Future work includes stronger network architectures, e.g., based on transformers
and with a separately trained post-net.
It also seems compelling to combine neural HMMs with powerful distribution families such as normalising flows, 
either replacing the Gaussian assumption (as done for non-neural HMMs in \cite{ghosh2021normalizing}) or as a probabilistic post-net like in \cite{kim2020glow}.
This may allow the naturalness of sampled speech to surpass that of deterministic output generation.

\bibliographystyle{IEEEbib}
\bibliography{refs_abbrev_2}

\end{document}